\documentclass[useAMS,usenatbib]{mn2e}

\usepackage{amsmath, amssymb}
\usepackage[dvips]{graphicx}
\usepackage{epsfig}

\newcommand{\grad}{\mathbf{\nabla}\,}
\newcommand{\rot}{\mathrm{\mathbf{rot}}\,}
\newcommand{\divg}{{\rm div}\,}

\newcommand{\rlight}{r_{\rm L}}

\newcommand{\er}{\vec{e}_{\rm r}}
\newcommand{\etheta}{\vec{e}_\vartheta}
\newcommand{\ephi}{\vec{e}_\varphi}

\newcommand{\aap}{A\&A}
\newcommand{\mnras}{MNRAS}

\newcommand{\apj}{ApJ}
\newcommand{\apjl}{ApJL}

\newcommand{\pre}{Physical Review E}

\title[The pulsar force-free magnetosphere linked to its striped
wind]{The pulsar force-free magnetosphere linked to its striped wind:
  time-dependent pseudo-spectral simulations} \author[J.  P\'etri]{J.
  P\'etri$^{1}$\thanks{E-mail:
    jerome.petri@astro.unistra.fr}\\
  $^{1}$Observatoire Astronomique de Strasbourg, Universit\'e de
  Strasbourg, CNRS, UMR 7550, 11 rue de l'Universit\'e, 67000
  Strasbourg, France.}

\begin{document}

\date{Accepted . Received ; in original form }

\pagerange{\pageref{firstpage}--\pageref{lastpage}} \pubyear{2011}

\maketitle

\label{firstpage}

\begin{abstract}
  Pulsar activity and its related radiation mechanism are usually
  explained by invoking some plasma processes occurring inside the
  magnetosphere, be it polar caps, outer/slot gaps or in the
  transition region between the quasi-static magnetic dipole regime
  and the wave zone, like the striped wind. Despite many detailed
  local investigations, the global electrodynamics around those
  neutron stars remains poorly described with only little quantitative
  studies on the largest scales, i.e. of several light-cylinder
  radii~$\rlight$.  Better understanding of these compact objects
  requires a deep and accurate knowledge of their immediate
  electromagnetic surrounding within the magnetosphere and its link to
  the relativistic pulsar wind.  This is compulsory to make any
  reliable predictions about the whole electric circuit, the energy
  losses, the sites of particle acceleration and the possibly
  associated emission mechanisms.

  The aim of this work is to present accurate solutions to the nearly
  stationary force-free pulsar magnetosphere and its link to the
  striped wind, for various spin periods and arbitrary inclination.
  To this end, the time-dependent Maxwell equations are solved in
  spherical geometry in the force-free approximation using a vector
  spherical harmonic expansion of the electromagnetic field. An exact
  analytical enforcement of the divergenceless of the magnetic part is
  obtained by a projection method.  Special care has been given to
  design an algorithm able to look deeply into the magnetosphere with
  physically realistic ratios of stellar~$R_*$ to
  light-cylinder~$\rlight$ radius.  However, currently available
  computational resources allows us only to set $R_*/\rlight =
  10^{-1}$ corresponding to pulsars with period $2$~ms. The spherical
  geometry permits a proper and mathematically well-posed imposition
  of self-consistent physical boundary conditions on the stellar
  crust. We checked our code against several analytical solutions,
  like the Deutsch vacuum rotator solution and the Michel monopole
  field. We also retrieve energy losses comparable to the
  magneto-dipole radiation formula and consistent with previous
  similar works.  Finally, for arbitrary obliquity, we give an
  expression for the total electric charge of the system. It does not
  vanish except for the perpendicular rotator. This is due to the
  often ignored point charge located at the centre of the neutron
  star. It is questionable if such solutions with huge electric
  charges could exist in reality except for configurations close to an
  orthogonal rotator.  The charge spread over the stellar crust is not
  a tunable parameter as is often hypothesized.
\end{abstract}

\begin{keywords}
  Pulsars: general - magnetic fields - MHD - plasmas - methods:
  numerical
\end{keywords}

\section{INTRODUCTION}

The problem of energy losses, i.e. particle acceleration and
radiation, from an active pulsar is closely related to the electric
current circulation within its magnetosphere, the geometry of close
and open magnetic field lines, the processes of gap formation and the
way the current is closed. It has been argued that this current flows
outside the light-cylinder, where very strong dissipation arises and
manifests itself as high-energy radiation.  Because the
electromagnetic field dominates the dynamics at least close to the
neutron star surface, low or even vanishing particle inertia and zero
pressure is often assumed.  This corresponds to a zero-th order
approximate solution to the pulsar magnetosphere and refered as
relativistic force-free electrodynamics or more properly
electromagnetodynamics. Steady axisymmetric magnetospheres have been
treated on an analytical basis for charge-separated plasma by
\cite{1974MNRAS.167..457O} and then extended to normal plasma by the
same author \citep{1975MNRAS.170...81O}. It is well known that the
force-free magnetic field in the far zone well outside the
light-cylinder should create a current sheet propagating outwards like
a vacuum wave \citep{1977MNRAS.180..125B}. The force-free
approximations has been applied to magnetospheres of compact objects,
especially to those of pulsars, known as the pulsar equation, for the
axisymmetric rotator \citep{1973ApJ...180..207M, 1973ApJ...182..951S,
  1974ApJ...187..359E} with attempts to solve it numerically
\citep{2006astro.ph..4364G}. An exact analytical solution has been
found for the special case of a magnetic monopole
\citep{1973ApJ...180L.133M}.  Although not realistic close to the
neutron star, it gives some insight into the far field solution,
reaching asymptotically a purely radial structure.  Many authors
investigated the aligned rotator because axisymmetry and time
independence lead to some drastic simplifications, although this does
not really correspond to a pulsar \citep{1999ApJ...511..351C,
  2006MNRAS.368.1055T, 2006MNRAS.367...19K, 2006MNRAS.368L..30M}.
Therefore, several authors tried to construct a force-free pulsar
magnetosphere by direct time-dependent simulations of Maxwell
equations in the general oblique case \citep{2006ApJ...648L..51S}.
Difficulties with outer boundary conditions due to reflection, cured
by perfectly matching layers \citep{2009A&A...496..495K} improved this
algorithm by implementing some absorbing outer boundary conditions.
However, the large ratio between stellar to light-cylinder radius
$R_*/\rlight = 0.2$ used in these previous works are not realistic for
observed pulsar periods (this would correspond to a 1~ms pulsar).
Moreover, the system has to reach a stationary state after some
relaxation time that needs several rotation periods and which is
difficult to control with boundary conditions that are not strictly
non-reflecting. However, this force-free approach was recently
extended by including resistivity \citep{2012ApJ...746...60L}.

Finding analytical or semi-analytical solutions is possible only in a
very few cases. For instance, \cite{1999MNRAS.309..388M} solved the
equations for the perpendicular rotator without aligned currents.
Therefore the linearity of the problem allowed them to employ Fourier
techniques for expanding the solution.

However, more realistic models should leave the force-free
description. Indeed \cite{2009MNRAS.398..271K} presented a two-fluid
solution to the aligned rotator.  Nevertheless, alternative models
with non-neutral fully charged separated magnetospheres containing
huge vacuum gaps have also been proposed and known as electrospheric
models. The early simulations of \cite{1985MNRAS.213P..43K} for the
aligned rotator have been improved by \cite{2002A&A...384..414P} and
extended to oblique rotators \citep{2009ApJ...690...13M}.

Back to the force-free simulations discussed above, some drawbacks
should be pointed out. First, using a Cartesian coordinate system
renders it painful to impose proper and convincing inner boundary
conditions on the stellar surface, to a good approximation depicted as
a perfect sphere.  Investigating the polar cap configuration becomes
therefore a difficult task. On the outer boundary, reflecting outgoing
waves can hardly be avoided as explained. Therefore, relaxation to a
stationary state becomes difficult to recognize. Probably the
strongest flaw is that the full expression for the force-free current
density is not taken into account, the part along the magnetic field
line being dropped because of its intricated expression including
spatial derivatives, tricky to handle with finite difference schemes.
\cite{2011arXiv1110.6669P} improved the situation by implementing a
pseudo-spectral code in spherical geometry for the axisymmetric pulsar
magnetosphere. Unfortunately, the code presented in this paper does
not strictly, or at least numerically, satisfy the divergenceless of
the magnetic field,~$\divg\mathbf{B}=0$, for instance by an explicit
imposition of this condition. Spurious magnetic monopoles could
locally be present when evolving Maxwell equations in time. Moreover
it seems that regularity conditions at the poles were overlooked and
are not ensured. This is probably due to inadequate expansion
techniques of a vector field viewed as a set of scalar fields for each
of its components.

On a more fundamental side, \cite{1997PhRvE..56.2181U} presented a
formalism introducing Euler potential to solve the time-dependent
force-free equations with special emphasize to symmetries
\citep{1997PhRvE..56.2198U}.  \cite{2003ApJ...583..842P} studied the
waves allowed by the force-free approximation and found that the
Alfven and fast modes behave similarly to MHD waves in the limit of
vanishing particle inertia.

In this paper, we demonstrate how to overcome many of the
aforementioned limitations by a proper and efficient expansion of a
general vector field onto vector spherical harmonics, ensuring
regularity and smoothness across the poles. We would like to stress
that the simulations presented here were performed on a single
processor for several hours to days or weeks and relatively modest
grids of resolution~$N_r \times N_\vartheta \times N_\varphi = 257
\times 32 \times 64$ at most. This is in clear contrast with finite
difference simulations employing grids as huge as $1000^3$ therefore
inevitably requiring parallelization and/or supercomputers. We present
full force-free pulsar magnetosphere solutions in nearly stationary
state using spherical polar coordinates. The outline of this paper is
as follows.  The time-dependent force-free model is exposed and
compared to a situation where the regime is strictly
stationary,~\S\ref{sec:Modele}. The pseudo-spectral method of solution
is presented in~\S\ref{sec:Algorithm}. The code is then tested and
validated against exact analytical solutions,~\S\ref{sec:Test}. Next,
the magnetospheric geometry and properties are described in depth
in~\S\ref{sec:Results}, providing the link to the base of the striped
wind, region believed to produce the high-energy pulsed emission
\citep{2009A&A...503...13P, 2010ApJ...715.1282B, 2011MNRAS.412.1870P}.
The magnetic topology is presented for an aligned, an oblique, and an
orthogonal rotator, energy losses versus obliquity and total electric
charge.  Conclusions and possible extensions are drawn
in~\S\ref{sec:Conclusion}.

\section{THE MODEL}
\label{sec:Modele}

In this section, we briefly recall the time-dependent Maxwell
equations in a force-free regime.  Next we address the problem of
stationary force-free equations for an oblique rotator and demonstrate
that the magnetic field is the main unknown from which all other
meaningful quantities are immediately computed. Actually the problem
is reduced to two kind of scalar fields by a vector spherical harmonic
expansion. The equations for the stationary state serve as an a
posteriori check for the time-dependent simulations to their closeness
to stationarity.

\subsection{Time-dependent Maxwell equations}

The time-dependent Maxwell equations in a flat space-time read
\begin{eqnarray}
  \label{eq:MaxwellTempsB}
  \partial_t \mathbf{B} & = & - \rot \mathbf{E} \\  
  \label{eq:MaxwellTempsE}
  \partial_t \mathbf{E} & = & c^2 \, \rot \mathbf{B} - \frac{\mathbf{j}}{\varepsilon_0}
\end{eqnarray}
supplemented with the two initial conditions
\begin{eqnarray}
  \label{eq:MaxwellInit}
  \divg \mathbf{B} & = & 0 \\  
  \divg \mathbf{E} & = & \frac{\rho_{\rm e}}{\varepsilon_0}
\end{eqnarray}
In a magnetically dominated relativistic flow such as those expected
in pulsar magnetospheres, the force-free approximation implies
negligible particle inertia as well as negligible thermal
pressure. The Lorentz force acting on a plasma fluid element must
therefore vanish, i.e.
\begin{equation}
  \label{eq:Force_Free}
  \rho_{\rm e} \, \mathbf{E} + \mathbf{j} \wedge \mathbf{B} = \mathbf 0
\end{equation}
the charge density being deduced from Maxwell-Gauss equation as
\begin{equation}
  \label{eq:Charge}
  \rho_{\rm e} \equiv \varepsilon_0 \, {\rm div} \, \mathbf E
\end{equation}
This defines the particle density number with respect to the electric
field and not conversely. Thus we assume that there is a source of
charge if necessary to maintain the force-free electromagnetic field.

From the force-free condition and the set of Maxwell equations, the
current density is found to be \citep{2006astro.ph..4364G}
\begin{equation}
  \label{eq:J_Ideal}
  \mathbf{j} = \rho_{\rm e} \, \frac{\mathbf{E}\wedge \mathbf{B}}{B^2} + 
  \frac{\mathbf{B} \cdot \rot \mathbf{B} / \mu_0 - 
  \varepsilon_0 \, \mathbf{E} \cdot \rot \mathbf{E}}{B^2} \, \mathbf{B}
\end{equation}
The first term represents the contribution from convection of charges
due to the electric drift, whereas the second contribution represents
the current sustained along magnetic field lines. This expression
holds only for magnetically dominated flows. Regions where this
condition would not be fulfilled could in principle form behind the
light-cylinder and should be avoid. It is difficult to design an
elliptic solver passing through the critical points of the flow and
meanwhile imposing the conditions for the force-free approximation. We
tried different iterative algorithms with several prescriptions behind
the light-cylinder but did not get highly accurate solutions as we
would expect from spectral methods. We therefore decided to switch
back to a time dependent formulation of the problem as exposed in the
previous lines. These time-dependent Maxwell equations
(\ref{eq:MaxwellTempsB}) and~(\ref{eq:MaxwellTempsE}) are integrated
with standard explicit schemes as will be explained in section
\ref{sec:Algorithm}.

\subsection{Towards the stationary state}

It is difficult to decide whether or not the system has reached a
steady state at the end of the run. This is especially true for non
axisymmetric magnetospheres for which all fields remain time-dependent
in the observer frame. It is therefore useful to get a more
quantitative criterion for stationary as described in this paragraph.
In a quasi-stationary state, the neutron star drags its magnetosphere
and its wind into a rigidly corotating body with constant angular
speed~$\Omega$. Mathematically speaking, this is written as an
equivalence between partial time derivative and partial azimuthal
derivative such that for any vector field~$\mathbf{V}$
\citep{2005ApJ...630..454M}
\begin{equation}
  \label{eq:Stationarite}
  \partial_t \mathbf V = - \Omega \, \partial_\varphi \mathbf V + \mathbf
  \Omega \wedge \mathbf V = \rot (\mathbf \beta \wedge \mathbf V) - \mathbf \beta
  \, {\rm div} \, \mathbf V 
\end{equation}
where we introduced spherical polar coordinates denoted by
$(t,r,\vartheta,\varphi)$ and $\mathbf\beta = \mathbf \Omega \wedge
\mathbf r$ is the corotation velocity.  Note that the partial
derivative with respect to $\varphi$
\begin{equation}
  \partial_\varphi \mathbf V = \partial_\varphi (V^r \, \er + V^\vartheta \, \etheta + V^\varphi \, \ephi )
\end{equation}
includes the orthonormal basis decomposition~$(\er, \etheta, \ephi)$.
Therefore this basis will also suffer some transformations during the
differentiation process. This explains the presence of the cross
product $\mathbf \Omega \wedge \mathbf V$ in the second term in the
middle part of equation~(\ref{eq:Stationarite}). Stated more simply,
it means that we only need to take partial derivatives of the
components of the vector $\mathbf V$ and not of the vector itself
(both being different due to the curvilinear coordinate system).
Thanks to this stationarity assumption, Maxwell-Amp\`ere equation is
integrated into
\begin{equation}
  \label{eq:Champ_Electrique}
  \mathbf E = - \mathbf \beta \wedge \mathbf B - \grad \Psi
\end{equation}
$\Psi$ being the electric potential as measured in the stellar
corotating frame. 
It is well known that force-free electrodynamics does not allow
particle acceleration along magnetic field lines because $\mathbf
E\cdot\mathbf B = \mathbf{B} \cdot\grad \Psi= 0$.  This also implies
that magnetic field lines are equipotentials for~$\Psi$. Moreover,
recalling that in the stellar interior, assumed to be a perfect
conductor, we have
\begin{equation}
  \label{eq:Champ_Electrique_Ideal}
  \mathbf E = - \mathbf \beta \wedge \mathbf B
\end{equation}
implying $\grad \Psi = \mathbf 0$. If there is no gap within the
magnetosphere, this assertion must hold in whole space. Therefore
Eq.~(\ref{eq:Champ_Electrique_Ideal}) remains valid everywhere, in the
neutron star as well as in its magnetosphere. As a consequence, the
electric field is just an unessential auxiliary unknown. More
importantly, this expression for the electric field automatically
satisfies the force-free regime, i.e. $\mathbf E \cdot \mathbf B = 0$.
Moreover this implies that at least within the light-cylinder, the
flow remains magnetically dominated i.e. $E<c\,B$.  To avoid
unphysical solutions it is compulsory to add a significant toroidal
component of the magnetic field outside the light-cylinder in order to
reduce the electric drift speed. Nothing forbids us to extend the
investigation well behind the light-cylinder but we were not able so
far to find an efficient and self-consistent way to enforce this
condition $E<c\,B$ for the boundary value problem implied by the
stationary magnetosphere equations. On a more physical ground, in this
region, we must either introduce an effective dissipation mechanism or
allow for particle acceleration, abandoning the force-free assumption.
A numerical trick to enforce magnetically dominated flows everywhere
consists to reset the electric field to values less than $c\,B$ in
appropriate regions.  We are then naturally led to a kind of
relaxation process that can be efficiently cast into a time-dependent
problem. The algorithm would therefore be very similar to the true
time-dependent Maxwell equations so we decided eventually to switch
back to time-dependent simulations of the magnetosphere.

It is readily seen that the charge density Eq.(\ref{eq:Charge}), the
current density, Eq.(\ref{eq:J_Ideal}) and the electric field
Eq.(\ref{eq:Champ_Electrique_Ideal}) are all derived from the magnetic
field structure. The latter is the only relevant unknown in a
stationary force-free problem.

In order to investigate the closeness of the time-dependent solution
to a stationary state, we check a posteriori that $\mathbf E \approx -
\mathbf \beta \wedge \mathbf B$. Moreover, the stationary
Maxwell-Amp\`ere equation reads
\begin{equation}
  \label{eq:Onde_Magnetique}
  \rot \mathbf{B} = \mu_0 \, \mathbf{j} - \frac{\Omega}{c^2} \,
  \partial_\varphi \mathbf E + \frac{\mathbf \Omega}{c^2} \wedge \mathbf E
\end{equation}
which furnishes the second criterion to check the relevance of
relaxation to a stationary state.

\section{Numerical algorithm}
\label{sec:Algorithm}

Finding the structure of the stationary pulsar magnetosphere is
equivalent to relaxing the time-dependent Maxwell equations towards a
stationary state with vanishing partial time derivatives $\partial_t
\mathbf{E} = \partial_t \mathbf{B} = 0$ in the corotating frame or
satisfying Eq.(\ref{eq:Stationarite}) in the lab frame. To this end,
we developed a pseudo-spectral collocation method in space to compute
spatial derivatives and augmented by a standard explicit in time ODE
integration scheme. The real strength of our code is the use of
spherical geometry without coordinate singularity along the polar axis
and proper boundary conditions on the stellar surface thanks to the
vector spherical harmonic expansion exposed in many details in the
appendix~\ref{app:HSV}. Regularity and smoothness conditions are also
automatically satisfied by the vector fields.

\subsection{Vector expansion}

A clever expansion of these vector fields is the heart of the
code. Indeed, electric and magnetic fields are expanded into vector
spherical harmonics (VSH) according to
\begin{eqnarray}
  \label{eq:E_vhs}
  \mathbf{E} & = & \sum_{l=0}^\infty\sum_{m=-l}^l
  \left(E^r_{lm} \mathbf{Y}_{lm} + E^{(1)}_{lm}\mathbf{\Psi}_{lm}+
    E^{(2)}_{lm}\mathbf{\Phi}_{lm}\right) \\
  \label{eq:B_vhs}
  \mathbf{B} & = & \sum_{l=0}^\infty\sum_{m=-l}^l
  \left(B^r_{lm} \mathbf{Y}_{lm} + B^{(1)}_{lm} \mathbf{\Psi}_{lm}+
    B^{(2)}_{lm}\mathbf{\Phi}_{lm}\right)
\end{eqnarray}
Note that the series expansions contain a finite number of terms, each
of them being smooth. As a consequence, the associated vector fields
will also remain smooth in the whole computational domain. By
definition, it is impossible to rigorously catch a discontinuity with
spectral methods. The best we can do is to introduce artificially a
transition layer of negligible thickness. From the expansion
coefficients, the linear differential operators like $\divg$ and
$\rot$ can be easily computed in the coefficient space and then
inverted back to real space.

It is a special case of spectral methods, known to possess
intrinsically very low numerical dissipation and able to resolve sharp
boundary layers \citep{Boyd2001}. However, if discontinuities arise in
the solution, due to the Gibbs phenomenon, the convergence rate is
drastically altered.  This is cured by applying some filtering process
to avoid aliasing and to smear the solution, which is equivalent to
some very low dissipation and known as super spectral viscosity
\citep{Ma:1998:CSS:291524.291526}. Various kind of filters can be used
as described for instance in \cite{Canuto2006}.

Compared to previous works, our approach has severals advantages.
First, because the expression for the current density is quite
intricated, the field aligned component is usually dropped and
replaced by a prescription for dissipation. Here we implement the full
expression, convection and field aligned contribution, reducing
numerical dissipation to the lowest possible value. Second, outgoing
wave boundary conditions are handled exactly by a characteristic
compatibility method. Third, inner boundary conditions are prescribed
on the stellar surface, i.e on a sphere. We impose continuous electric
field components~$\{E_\vartheta,E_\varphi\}$ as well as a continuous
$B_r$ component. Thus the system is well posed and reduces to Deutsch
solution for a magnetosphere without current.  We now give some
details about the algorithm.

\subsection{Maintaining the force-free requirements and the
  divergenceless of $\mathbf{B}$}

During the time evolution of the electromagnetic field, the force-free
conditions will be violated soon or later. It is therefore necessary
to reinforce the $\mathbf{E}\cdot\mathbf{B}=0$ and $E<c\,B$ conditions
regularly. Consequently, at each time step, in a first stage, we
subtract the magnetic field aligned electric field component by
adjusting the new electric field~$\mathbf{E}'$ such that
\begin{equation}
  \label{eq:EB0}
  \mathbf{E}' = \mathbf{E} - \frac{\mathbf{E}\cdot\mathbf{B}}{B^2} \, \mathbf{B}
\end{equation}
This new electric field need not satisfy the requirement $E'<c\,B$. In
a second stage, we therefore reduce the electric field~$\mathbf{E}'$ by a factor
$E'/c\,B$ to finally get the corrected electric field as
\begin{equation}
  \label{eq:E_cor}
  \mathbf{E}_{\rm cor} = \mathbf{E}' \,\sqrt{\frac{c^2\,B^2}{E'^2}}
\end{equation}
The electric field is updated with this corrected value where
necessary.  Next, to insure a divergenceless magnetic field, at each
time step, we project the vector $\mathbf{B}$ onto the subspace
defined by $\divg \mathbf{B} = 0$ simply by applying a forward
transform to the coefficients $\{f^B_{lm}(r,t),g^B_{lm}(r,t)\}$
according to Eq.~(\ref{eq:Decomposition_HSV_div_0}) followed by a
backward transform to the vector field $\mathbf{B}$, which is
analytically divergenceless by definition. This procedure removes the
longitudinal part of the magnetic field.

\subsection{Exact boundary conditions}

The spherical geometry of our code allows an exact enforcement of the
boundary conditions on the star. Moreover, (pseudo)-spectral methods
reflect perfectly the underlying analytical problem and therefore only
requires the same boundary conditions as the original mathematical
problem. There is no need for extra boundary conditions as would be
the case for finite difference methods. This would make the problem
ill-posed and the code unstable. Finite difference algorithms being
much more dissipative, they will damp these unstable solutions!

The correct jump conditions at the inner boundary are, continuity of
the normal component of the magnetic field~$B_r$ and continuity of the
tangential component of the electric
field~$\{E_\vartheta,E_\varphi\}$. Explicitly, we have
\begin{eqnarray}
  \label{eq:CLimites}
  B_r(t,R_1,\vartheta,\varphi) & = & B_r^0(t,\vartheta,\varphi) \\
  E_\vartheta(t,R_1,\vartheta,\varphi) & = & - R_1 \, \Omega \, \sin\vartheta \, B_r^0(t,\vartheta,\varphi) \\
  E_\varphi(t,R_1,\vartheta,\varphi) & = & 0
\end{eqnarray}
Note that the continuity of $B_r$ supplemented with the condition
Eq.~(\ref{eq:Champ_Electrique_Ideal}) automatically implies the right
boundary treatment of the electric field.
$B_r^0(t,\vartheta,\varphi)$ represents the possibly time-dependent
radial magnetic field imposed by the star (monopole, split monopole,
aligned or oblique dipole).

As a corollary of these boundary conditions, it is impossible and even
not self-consistent to force the surface charge density to vanish on
the neutron star crust. This quantity will be derived a posteriori
from the results of the simulations.

The outer boundary conditions are more subtle to handle. Ideally, we
would like outgoing wave conditions in order to prevent reflection
from the artificial outer boundary. The appropriate technique is
called Characteristic Compatibility Method (CCM) and described in
\cite{Canuto2007}. We have implemented this technique in spherical
geometry. Indeed, the exact radially propagating characteristics are
known and given by
\begin{eqnarray}
  \label{eq:CCM1}
  E_\vartheta \pm c\, B_\varphi & ; & E_\varphi \pm c\, B_\vartheta
\end{eqnarray}
within an unimportant factor $r$. In order to forbid ingoing wave we ensure
\begin{eqnarray}
  \label{eq:CCM2}
  E_\vartheta - c\, B_\varphi & = & 0 \\
  \label{eq:CCM3}
  E_\varphi + c\, B_\vartheta & = & 0
\end{eqnarray}
whereas the other two characteristics are found by
\begin{eqnarray}
  \label{eq:CCM4}
  E_\vartheta + c\, B_\varphi & = & E_\vartheta^{\rm PDE} + c\, B_\varphi^{\rm PDE} \\
  \label{eq:CCM5}
  E_\varphi - c\, B_\vartheta & = & E_\varphi^{\rm PDE} - c\, B_\vartheta^{\rm PDE}
\end{eqnarray}
the superscript $^{\rm PDE}$ denoting the values of the
electromagnetic field obtained by straightforward time advancing
without care of any boundary condition. The new corrected values are
deduced from Eq.~(\ref{eq:CCM2})-(\ref{eq:CCM5}). We expect to reach
stationarity in the whole computational domain within a reasonable
time span.

\subsection{Dealiasing and filtering}

The source term of Maxwell equations represented by the force-free
current density Eq.~(\ref{eq:J_Ideal}) introduces a strong
non-linearity in the problem and therefore aliasing effect for
unresolved spatial scales. In principle, the numerator being a cubic
product of unknown functions could be cured by zero-padding techniques
\citep{Canuto2006}, but the denominator renders this task impossible.
Dealiasing of a quotient of two functions can efficiently be achieved
by inverting a linear system for one dimensional problem
\citep{Debusschere:2005:NCU:1039891.1039930}.  Unfortunately, this is
intractable for our three-dimensional spherical problem. We prefer to
use filtering by smoothing the electromagnetic field at each time step
after reinforcing the force-free conditions. We use an eighth order
exponential filter in all directions with
\begin{equation}
  \label{eq:Filtre}
  \sigma(\eta) = \textrm{e}^{-\alpha\,\eta^\beta}
\end{equation}
$\eta$ ranges between 0 and 1. For instance, in the radial coordinate
$\eta=k/(N_r-1)$ for $k\in[0..N_r-1]$, $k$ being the index of the
coefficient $c_k$ in the Chebyshev expansion
$f(x)=\sum_{k=0}^{N_r-1} c_k\,T_k(x)$. We tried several parameter sets, a good
compromise between accuracy and stability is $\alpha=10$ and
$\beta=8$.

\subsection{Time integration}

Pseudo-spectral methods aim at replacing a set of partial differential
equations (PDE) by a set of ordinary differential equations (ODE) for
the unknown collocation points or spectral
coefficients. Schematically, we write it as
\begin{equation}
  \label{eq:ODE}
  \frac{d\mathbf{u}}{dt} = f(t,\mathbf{u})
\end{equation}
with appropriate initial and boundary conditions. $\mathbf{u}$
represents the vector of unknown functions either evaluated at the
collocation points or simply the spectral coefficients.

Several methods for integrating the time evolution of those ODE are at
hand. It is worth mentioning implicit versus explicit schemes and
single versus multi-step marching methods. Runge-Kutta schemes remain very
popular as a single step scheme, but their main flaw is the evaluation
of the RHS very often. For spectral methods, Adam-Bashforth (explicit
scheme) and Adam-Moulton (implicit scheme), known as multi-step
integrators, are also widely employed. Their advantage is that no
extra cost is needed to compute the right hand side.

Although many time marching algorithms could be justified, we decided
to use a third order Adam-Bashforth scheme advancing a vector of
unknown functions~$\mathbf{u}$ as
\begin{equation}
  \label{eq:AB3}
  \mathbf{u}^{n+1} = \mathbf{u}^{n} + 
  \Delta t \, \left( \frac{23}{12} \mathbf{f}^{n} - 
    \frac{16}{12} \mathbf{f}^{n-1} + \frac{5}{12} \mathbf{f}^{n-2} \right)
\end{equation}
where $\Delta t$ is the \textit{fixed} time step subject to some
stability restrictions, $\mathbf{u}^{n} = \mathbf{u}(n\,\Delta t)$ and
$\mathbf{f}^{n} = \mathbf{f}(n\,\Delta t,\mathbf{u}(n\,\Delta t))$.

Let us say a few words about the time step restriction. For the
time-marching scheme to remain stable, $\Delta t$ should remain in the
stability region defined by the inequality $v_{\rm max}\,\Delta t\leq
\alpha \, \Delta x_{\rm min}$ where $v_{\rm max}$ is the maximum speed
of the waves, here equal to the speed of light~$c$, $\alpha$ a
constant close to unity depending on the integration scheme and
$\Delta x_{\rm min}$ the minimum grid spacing between two points. In
our spherical computation domain, the severe constraint comes from the
inner part because
\begin{equation}
  \label{eq:dxmin}
  \Delta x_{\rm min} = \textrm{Min}(\Delta r, r\,\Delta\vartheta, r\,\sin\vartheta\,\Delta\varphi)
\end{equation}
It is usually claimed that the non-uniform grid introduced by the
Chebyshev collocation points, which are very dense at the boundaries,
are the most critical conditions because~$\Delta r =
O(N_r^{-2})$. While this is true for Cartesian coordinates, we find
actually that the grid spacing along the polar axis imposes an even
stronger limit on the time step because of the presence of the factor
$r\,\sin\vartheta$ in front of $\Delta\varphi$, depending on the
resolution adopted in the radial and transverse directions. This
explains why we do not tried a mapping of the Chebyshev collocation
points by the popular Kosloff/Tal-Ezer grid
\citep{1993JCoPh.104..457K}. Such mapping unfortunately kills the
spectral convergence properties of the methods, degrading it only to a
second order scheme \citep{Boyd2001}.

\section{TEST}
\label{sec:Test}

Before handling the general oblique rotator, we check our algorithm
against some well known analytical solutions. The starting point is
the Deutsch vacuum field solution for a perpendicular rotator. Then,
we investigate the monopole solutions with a central monopole magnetic
field.

In the remaining of the paper, we adopt the following normalization:
the magnetic field strength at the light cylinder in the equatorial
plane is set to unity, $B_{\rm L}=1$, as well as the stellar angular
velocity and the speed of light $\Omega=c=1$, therefore $\rlight=1$.

\subsection{The Deutsch solution}

The vacuum electromagnetic field of an oblique rotator is known
analytically since the work by~\cite{1955AnAp...18....1D}. Because
Maxwell equations are linear in vacuum space, solutions are found
through complex analysis. Assuming a star of radius $R_*$ and surface
magnetic field of $B_*$ with rotation speed $\Omega_*$ and
obliquity~$\chi$, the complex electromagnetic field is easily computed
and given in the appendix~\ref{app:Deutsch}.

We start the simulation with the perpendicular static dipole magnetic
field ($\chi=90\degr$) and zero electric field outside the star,
except for the crust where we enforce the inner boundary condition
with corotating electric field, see Sec.~\ref{sec:Algorithm}.  We
performed simulations with different radial extensions of the
computational domain and tried several resolutions. For the largest
radial dimensions, finer grids in radius are required, therefore
increasing the number of radial collocation points~ $N_r$. Typically,
for a domain $r\in[0.2,2]$, a resolution of $N_r \times N_\vartheta
\times N_\varphi = 33\times16\times32$ is sufficient to get accurate
solutions. For the largest runs with $r\in[0.1,10]$ we found that a
minimum resolution of $N_r \times N_\vartheta \times N_\varphi =
257\times16\times32$ was necessary.

\begin{figure}
  \centering
  \includegraphics[width=0.45\textwidth]{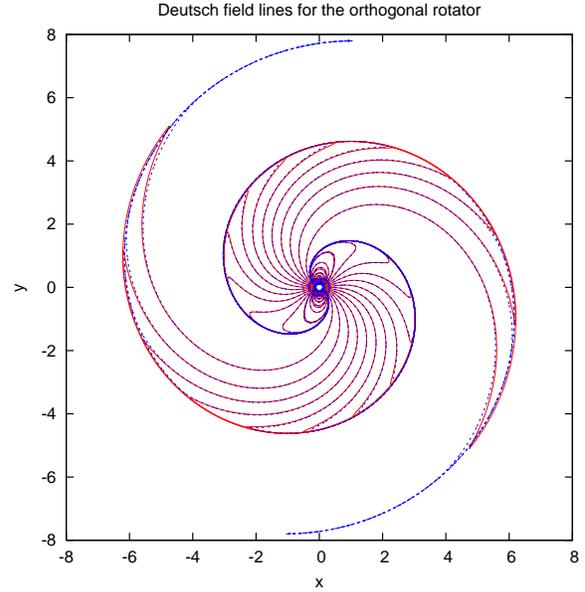}
  \caption{Magnetic field lines of the perpendicular Deutsch field in
    the equatorial plane. The exact analytical solution (solid red
    lines) is compared to the time-dependent simulation (dashed blue
    lines). The overlapping is excellent in the whole region where the
    relaxation to a quasi-stationary state is nearly achieved.}
  \label{fig:Deutsch}
\end{figure}

\begin{figure}
  \centering
  \includegraphics[width=0.45\textwidth]{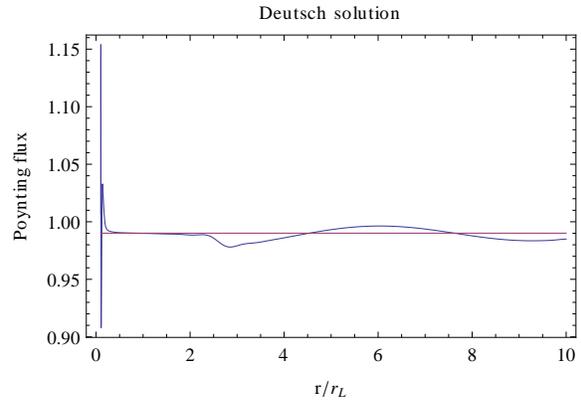}
  \caption{Poynting flux across the sphere of radius~$r$. The computed
    flux (blue line) is almost constant and equal to its analytical
    value (red line).  The solution settled down to the stationary
    Deutsch field even close to the outer boundary.}
  \label{fig:Deutsch_Poynting}
\end{figure}

We let the system evolve for two rotational period of the pulsar. The
final magnetic field line configuration in the equatorial plane is
shown as dashed blues curves in fig.~\ref{fig:Deutsch} and compared to
the analytical solution depicted as solid red curves. The numerical
solution is very close to the exact solution everywhere, they are
hardly distinguishable. The associated Poynting flux is shown in
fig.~\ref{fig:Deutsch_Poynting}. It is constant and equal to the
Deutsch value everywhere except close to the star. Nevertheless after
a thin boundary layer, the flux relaxes sharply to the exact
stationary value of Deutsch. In the outer parts of the domain, the
system evolved to a nearly steady state configuration and no
significant reflections have to be reported.  The Poynting flux is
normalized to the monopole solution flux, see
Eq.~(\ref{eq:L_mono}). It is a bit less than unity because it depends
on the ratio~$R_*/\rlight$ and tends to Eq.~(\ref{eq:L_mono}) when
$R_*/\rlight$ tends to zero.

This first test demonstrates the ability of our code to handle non
axisymmetric configuration with high accuracy and to relax the system
to an almost stationary state in a vacuum region.

\subsection{The monopole solution}

Next we tackle the problem of an axisymmetric force-free flow known as
the monopole field introduced by \cite{1973ApJ...180L.133M}.  We
recall that this monopole solution is given by
\begin{equation}
  \mathbf{B} = B_{\rm L} \, \frac{\rlight^2}{r^2} \, \er - B_{\rm L} \, \frac{\rlight}{r} \, \sin\vartheta \, \ephi
\end{equation}
In terms of a vector spherical harmonic (VSH) expansion, this magnetic
field is expressed as
\begin{equation}
 \mathbf{B} = B_{\rm L} \, \frac{\rlight^2}{r^2} \, \er + g_{10}^B(r) \, \mathbf{\Phi}_{10}  
\end{equation}
where
\begin{equation}
  \label{eq:gB10}
  g_{10}^B(r) = \sqrt{\frac{8\pi}{3}} \, B_L \, \frac{\rlight}{r}
\end{equation}
all other coefficients being equal to zero. See the
appendix~\ref{app:HSV} for a detailed exposure of the VSH, their
definition and useful properties.

\begin{figure}
  \centering
  \includegraphics[width=0.45\textwidth]{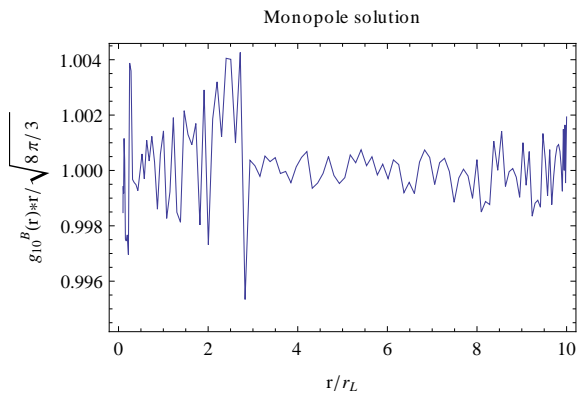}
  \caption{The relevant VSH coefficient~$g^B_{10}(r)$ for the monopole
    magnetic field expansion in the domain~$r/\rlight\in[0.1,10]$. For
    ease to compare, we applied a normalization factor of
    $r/\sqrt{8\pi/3}$. All other coefficients~$g^B_{lm}(r)$ with the
    same normalization factor are less than $10^{-6}$. Results are
    shown for a resolution of $N_r \times N_\vartheta \times N_\varphi
    = 257\times16\times32$.}
  \label{fig:Monopole}
\end{figure}

\begin{figure}
  \centering
  \includegraphics[width=0.45\textwidth]{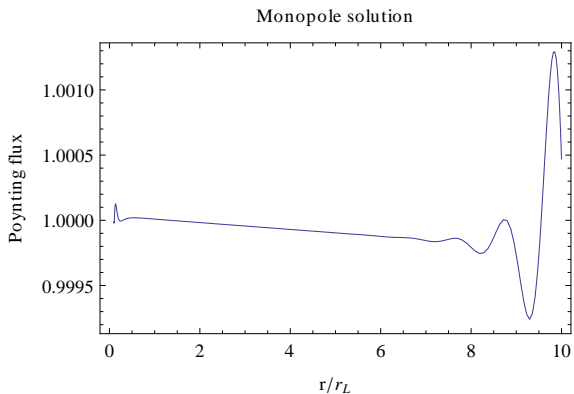}
  \caption{Poynting flux across the simulation sphere corresponding to
    the numerical solution of fig.~\ref{fig:Monopole}. It is
    approximately constant as expected from energy conservation
    arguments. The same resolution as in fig.~\ref{fig:Monopole}
    applies.}
  \label{fig:Monopole_Poynting}
\end{figure}

The run is started with a pure monopolar magnetic field and zero
electric field outside the star, as before. An Alfven wave is launched
from the stellar surface and quickly forces the system to a stationary
state. Here also, as expected, larger radial extension implies finer
grid in radius. Typically, for a domain $r\in[0.2,2]$, a resolution of
$N_r \times N_\vartheta \times N_\varphi = 33\times16\times32$ is
sufficient to get accurate solutions. For the largest runs with
$r\in[0.1,10]$ we found that a minimum resolution of $N_r \times
N_\vartheta \times N_\varphi = 257\times16\times32$ was necessary.
The relevant coefficient~$g^B_{10}(r)$ is shown in
fig.~\ref{fig:Monopole} for comparison with the analytical
solution. The numerical solution is very close to the exact solution
in the whole computational box. The associated Poynting flux,
normalized to the exact monopole value of
\begin{equation}
  \label{eq:L_mono}
  L_{\rm mono} = \frac{8\,\pi}{3\,\mu_0\,c^3} \, \Omega_*^4 \, B_*^2 \, R_*^6 =  \frac{8\,\pi}{3\,\mu_0\,c^3} \, \Omega_*^4 \, B_{\rm L}^2 \, \rlight^6
\end{equation}
is depicted in fig.~\ref{fig:Monopole_Poynting}. It is almost constant
and equal to the monopole value as expected. Note that there is no
spurious reflection in the outer regions, we are very close to a
steady state configuration.
We tried higher resolutions for instance $N_r \times N_\vartheta
\times N_\varphi = 257\times32\times64$ without significant
improvement of the computed solutions. All spatial scales have been
resolved with the coarsest grid. That such a low resolution is
sufficient is no surprise because the monopole solution does only
contain two coefficients: the monopole radial part $f^B_{00}(r) =
B_{\rm L} \, \rlight^2/r^2$ and the toroidal part $g^B_{10}(r)$, so
there exist at most one mode $l=1$.

\section{RESULTS}
\label{sec:Results}

We next look at the oblique pulsar magnetosphere, especially the
aligned rotator, studied several times in the past by several authors.
New results about the perpendicular rotator and its link to the
striped wind will also be shown.

\subsection{Simulation setup}

The rotating neutron star has a radius$~R_1$.  The computational
domain extends from this inner boundary~$R_1$ to an outer boundary
shell of radius $R_2 > R_1$. Note that the star does not belong to the
simulation box except for its surface in order to fix the inner
boundary conditions.  For concreteness, we start from
$R_1=0.2\,\rlight$ corresponding to a 1~ms pulsar down to
$R=0.1\,\rlight$ describing a 2~ms pulsar.  
%This helps to investigate the influence of the pulsar period onto the
%magnetic field geometry and energy losses.  
Likewise the extension of the simulation sphere is variable and equal
to $R_2=2\,\rlight$ up to $R_2=10\,\rlight$ to investigate the
beginning of the striped wind.  The inclination angle of the magnetic
moment with respect to the rotation axis was set to $\chi = \{ 0\degr,
30\degr, 60\degr, 90\degr\}$.

We performed several simulation sets with different filters, the most
common being Cesaro, raised cosinus and 2nd, 4th, 8th-order
exponential filter (see \cite{Canuto2006}) as well as different
spatial resolutions. We found that for spatially resolved
discretization, the algorithm converges quickly to a stationary
state. We emphasize that the solutions become independent of the
choice of a filter as it should be, given us confidence about the
convergence and reliability of our results.

\subsection{Magnetospheric structure}

\begin{figure}
  \centering
  \includegraphics[width=0.45\textwidth]{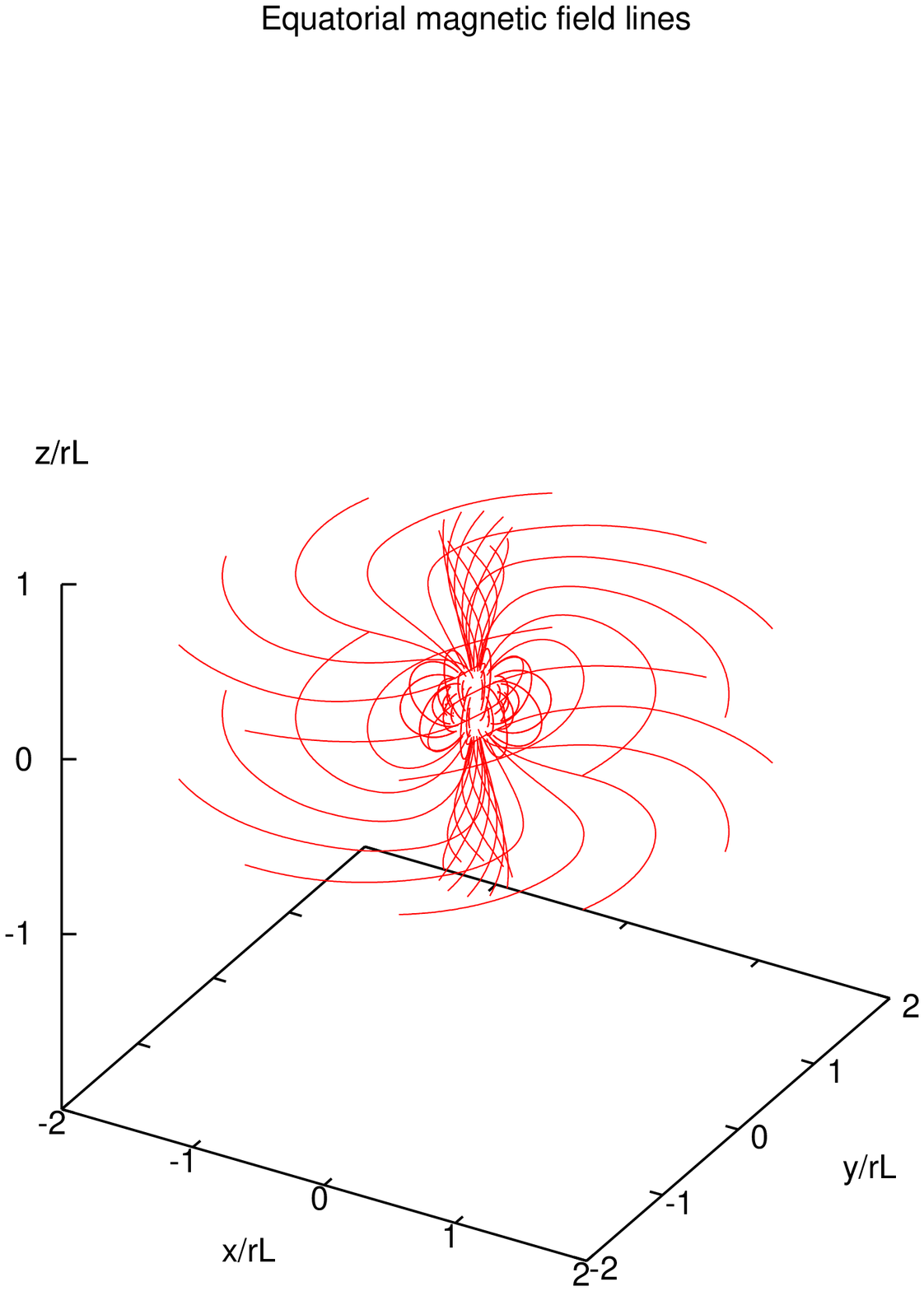}
  \caption{3D magnetic field lines for the aligned rotator. The
    inner and outer spherical boundaries are located respectively at
      $R_1=0.2\,\rlight$ and $R_2=2\,\rlight$.}
    \label{fig:Champ_Aligne}
  \end{figure}

\begin{figure}
    \centering
    \includegraphics[width=0.45\textwidth]{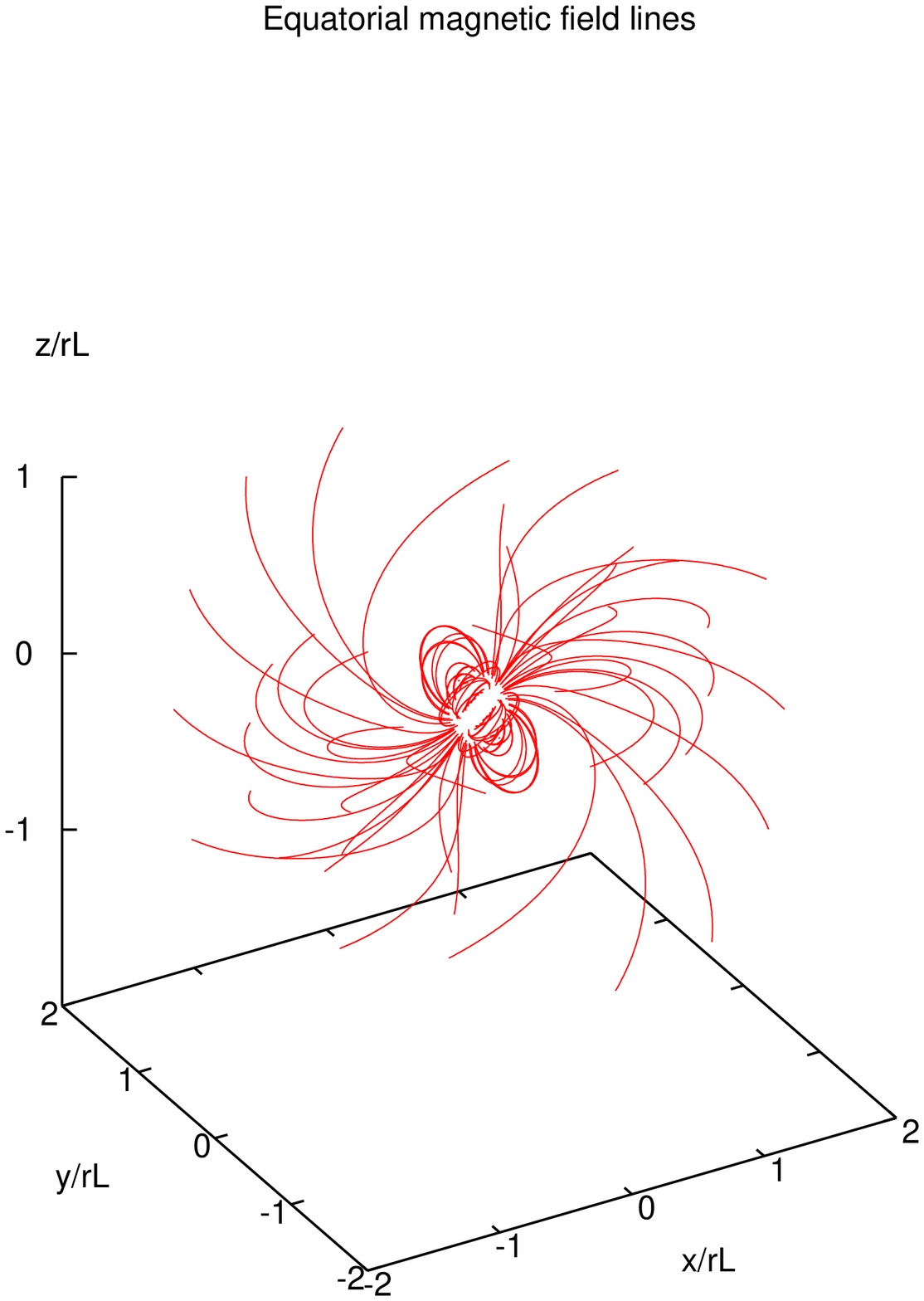}
    \caption{3D magnetic field lines for the inclined rotator with
      $\chi=60\degr$. The inner and outer spherical boundaries are located
      respectively at $R_1=0.2\,\rlight$ and $R_2=2\,\rlight$.}
    \label{fig:Champ_Oblique}
  \end{figure}

\begin{figure}
  \centering
  \includegraphics[width=0.45\textwidth]{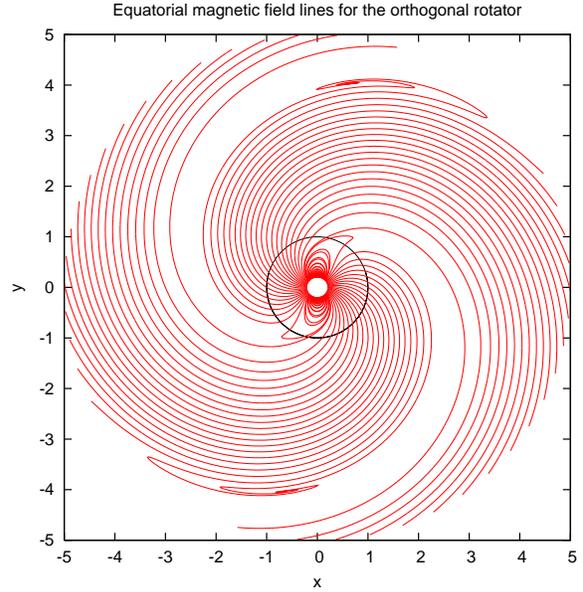}
  \caption{Equatorial magnetic field lines for the perpendicular
    rotator. The light cylinder is shown as a black circle. The inner
    and outer spherical boundaries are located respectively at
    $R_1=0.2\,\rlight$ and $R_2=5\,\rlight$.}
  \label{fig:Champ_Orthogonal}
\end{figure}

\begin{figure}
  \centering
  \includegraphics[width=0.45\textwidth]{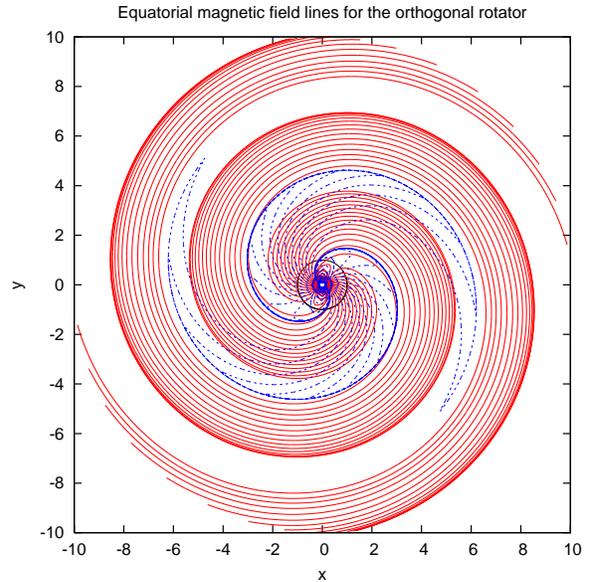}
  \caption{Equatorial magnetic field lines for the perpendicular
    force-free rotator (red solid line), compared to the vacuum
    Deutsch field (dashed blue line). The light cylinder is shown as a
    black circle. The inner and outer spherical boundaries are located
    respectively at $R_1=0.1\,\rlight$ and $R_2=10\,\rlight$. Note the
    similarity between both structures.}
  \label{fig:Champ_Orthogonal_2}
\end{figure}

\begin{figure}
  \centering
  \includegraphics[width=0.45\textwidth]{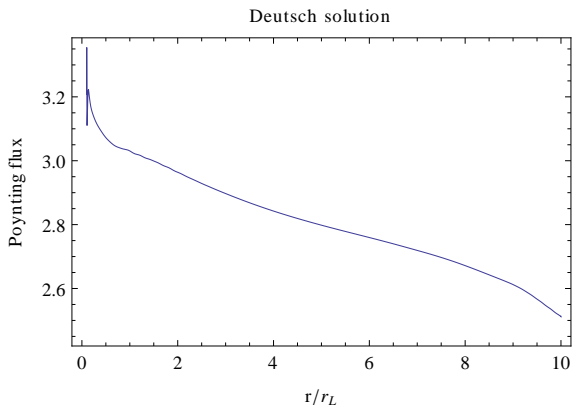}
  \caption{The Poynting flux across the whole simulation sphere for
    the orthogonal rotator of fig.~\ref{fig:Champ_Orthogonal_2}. The
    Poynting flux is nearly constant but artificial viscosity due to
    filtering slightly alters energy conservation outside the
    light-cylinder.}
  \label{fig:Champ_Orthogonal_3}
\end{figure}

First, let us have a look on the magnetic field geometry.  Let us
start with the special case of an aligned rotator.  The 3D magnetic
field lines are shown in fig.~\ref{fig:Champ_Aligne}. The inner
boundary is at $R_1 = 0.2 \, \rlight$ and the outer boundary at $R_2 =
2 \, \rlight$. Inside the light-cylinder, the geometry is close to the
static dipole without significant toroidal magnetic field component.
Along the polar caps and outside the light-cylinder, we recognize a
monopole structure with a significant toroidal component. The aligned
rotator, although being simple because axisymmetric and therefore more
tractable analytically and computationally, is of no relevance for
pulsar theory because it is not able to reproduce the pulsar
phenomenon. Inclined models are true pulsars.
  Thus, we give an example of oblique rotator as shown in
  fig.~\ref{fig:Champ_Oblique} with $\chi=60\degr$. The weakly
  disturbed static dipole regime inside the light-cylinder contrasts
  with the open field lines passing through the light-cylinder as is
  costumary for the force-free magnetosphere. Visualization and
  exploiting data is cumbersome for the full 3D geometry, so we do not
  discuss further this peculiar case. The orthogonal rotator with its
  inherent symmetry about the equatorial plane represents the best
  configuration to study a realistic pulsar magnetosphere.
  We show such rotator for which vizualisation is easier because some
  field lines are entirely contained in the equatorial plane. These
  lines are shown in fig.~\ref{fig:Champ_Orthogonal}. The inner
  boundary is at $R_1 = 0.2 \, \rlight$ and the outer boundary at $R_2
  = 5 \, \rlight$. The magnetic topology is reminiscent of the Deutsch
  field solution with some regions of strong deviation due to the
  magnetospheric current.  Inside the light-cylinder (black circle),
  the geometry is close to the static dipole whereas outside, the
  launch of the striped wind is seen with its spiral structure in
  regions of strong magnetic field gradient.
In order to compare the orthogonal force-free field with the vacuum
field, we did a large scale simulation with $R_1=0.1\,\rlight$ and
$R_2=10\,\rlight$ and shown in fig.~\ref{fig:Champ_Orthogonal_2}. The
shape of the spiral current sheet in the force-free striped wind
follows a geometric shape reminiscent to the Deutsch solution. The
locus of the current sheet in the striped wind according to the
rotational phase of the neutron star has strong implications on the
association between radio and high energy emission. The time lag
between the radio pulses and the gamma-ray pulses are direct
observational consequences and can be checked
\citep{2011MNRAS.412.1870P}.
Inspecting the Poynting flux dependence with respect to radius,
fig.~\ref{fig:Champ_Orthogonal_3}, we observed a tendency to dissipate
the electromagnetic flux well outside the light-cylinder. The employed
resolution should still be increased but must wait more computational
power.

\subsection{Energy losses and electric charge}

\begin{figure}
  \centering
  \includegraphics[width=0.45\textwidth]{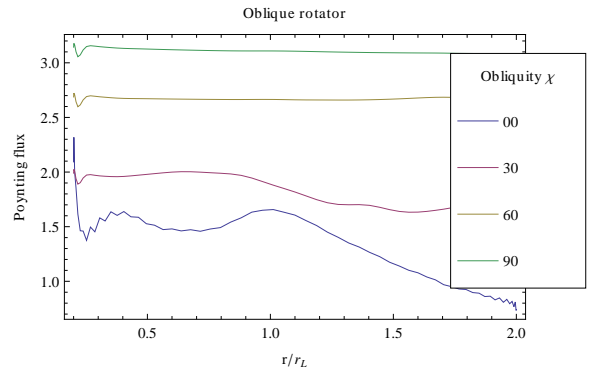}
  \caption{The Poynting flux across the whole simulation sphere as a
    function of the obliquity~$\chi$. The flux is nearly constant for
    the perpendicular rotator. Artificial dissipation due to filtering
    becomes significant for almost aligned rotators outside the
    light-cylinder.}
  \label{fig:Poynting}
\end{figure}

\begin{figure}
  \centering
  \includegraphics[width=0.4\textwidth]{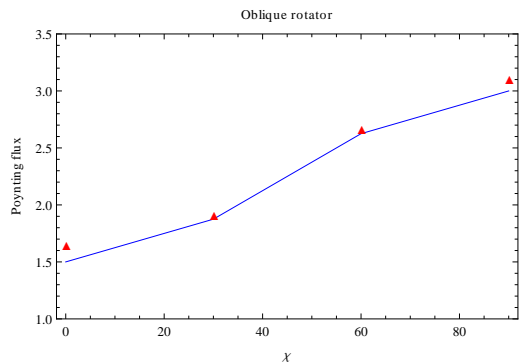}
  \caption{The Poynting flux across the sphere of radius~$\rlight$ as
    a function of the obliquity~$\chi$. Simulation results are
    depicted by red triangles and the fit is shown in blue solid
    line.}
  \label{fig:PoyntingFit}
\end{figure}

Another important issue is the spin-down energy of the pulsar by
magnetic braking. Thanks to the knowledge of the magnetospheric
structure, we can investigate the energetics of the pulsar
quantitatively.  Current wisdom assumes that the Poynting flux
escaping from the force-free magnetosphere is of the same order of
magnitude as the vacuum dipole rotator. We check this assertion by
plotting the Poynting flux across the simulation sphere as a function
of the obliquity with a simulation sphere extending from
$R_1=0.2\rlight$ to $R_2=2\rlight$, fig.~\ref{fig:Poynting},
normalized with respect to the magneto-dipole losses for an orthogonal
rotator expressed as
\begin{equation}
  \label{eq:Ldip}
  L_{\rm dip} = \frac{8\,\pi}{3} \, \frac{\Omega^4\,B^2\,R^6}{\mu_0\,c^3}.
\end{equation}
Within a factor of several unity, the luminosity is the same for
vacuum dipole and force-free magnetosphere, fig.~\ref{fig:Poynting}.
The flux across the light-cylinder, i.e. before artificial viscosity
due to filtering becomes significant can approximately be fitted with,
fig.~\ref{fig:PoyntingFit},
\begin{equation}
  \label{eq:L_dot}
  L_{\rm sp} \approx 1.5 \, L_{\rm dip} \, ( 1 + \sin^2\chi )
\end{equation}
Thus the orthogonal force-free magnetosphere radiates 3~times more
than the vacuum analog whereas the aligned magnetosphere radiates 50\%
of the vacuum perpendicular rotator. These results agree with the
functional form found by \cite{2006ApJ...648L..51S} although we took
into account the full expression for the current density.
Surprisingly, we found it easier to look for the perpendicular rotator
than for the aligned one. Indeed, in the latter case, the current
sheet in the equatorial plane is coarsely resolved by our current grid
resolution. According to the lower blue curve in
fig.~\ref{fig:Poynting}, non negligible artificial dissipation is
present and smears out the discontinuity. On the contrary, the strong
displacement current in the former case leads to more gentle current
sheet much better described by the coarse grid. Scrutiny of the upper
green curve in fig.~\ref{fig:Poynting} demonstrates indeed that the
Poynting flux remains nearly constant with radius. Energy is conserved
to good accuracy as is expected from the force-free evolution
equations \citep{2002MNRAS.336..759K}. In order to circumvent high
dissipation rate, the only efficient remedy would be to increase the
grid resolution. Unfortunately, for the moment we are unable to run
such high-resolution simulation on our computer. We plan to improve
our code by performing some optimizations and by using parallelization
techniques.

\begin{figure}
  \centering
  \includegraphics[width=0.45\textwidth]{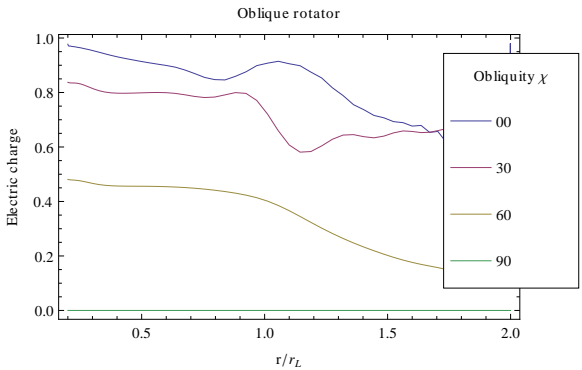}
  \caption{Total electric charge~$Q_{\rm tot}$ in the simulation
    sphere as a function of the obliquity~$\chi$ and normalized
    against the central point charge for the aligned rotator~$Q_{\rm
      c} / \cos\chi$. Simulation parameters are exactly the same as in
    fig.~\ref{fig:Poynting}.}
  \label{fig:Charge}
\end{figure}

\begin{figure}
  \centering
  \includegraphics[width=0.4\textwidth]{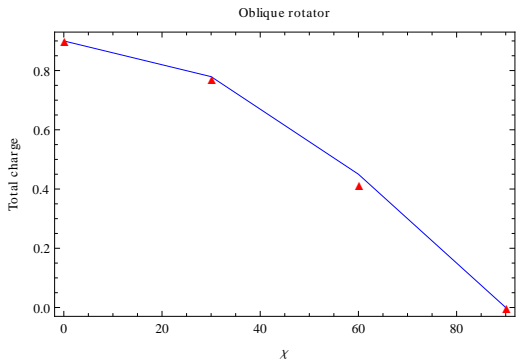}
  \caption{Total electric charge across the sphere of radius~$\rlight$
    as a function of the obliquity~$\chi$. Simulation results are
    depicted by red triangles and the fit is shown in blue solid
    line.}
  \label{fig:ChargeFit}
\end{figure}

Finally, the total electric charge enclosed by a spherical shell
centered at the pulsar location is computed by the electric flux
across it. It is too rarely stated that the total charge of the
neutron star and its magnetosphere is not equal to zero but very close
to the value of the central point charge $Q_{\rm c}$, located in the
middle of the star (see appendix~\ref{app:Deutsch} for the definition
of this charge).  Indeed, the total charge of the system with respect
to obliquity~$\chi$ is shown in fig.~\ref{fig:Charge}.  Only in the
special case of an orthogonal rotator is this charge exactly
zero. This is true for all ratio $R_*/\rlight$. The fit is simply given
by $ Q_{\rm ffe} \approx 0.9 \, Q_{\rm c}$, fig.~\ref{fig:ChargeFit}.
Here also, the dependence on $\cos\chi$ is reminiscent of the central
point charge~$Q_{\rm c}$.

\section{CONCLUSION}
\label{sec:Conclusion}

We solved the full set of time-dependent force-free equations for a
rotating oblique pulsar magnetosphere. We tested our pseudo-spectral
collocation code and demonstrated that accurate numerical solutions
can be found for smooth fields.  Interestingly, we found it easier to
compute the solution for the perpendicular rotator because the flaw of
any spectral method to catch discontinuities is harmless in that
particular geometry. It seems that the presence of a strong
displacement current facilitates the computation in the vicinity of
the current sheet. Unfortunately, the current version of our code does
not allow to perform very high resolution on a single processor in a
reasonable time to reduce the Gibbs phenomenon for non-smooth
configurations as those close to the aligned rotator. We plan to
improve the code by optimization and parallelization techniques in the
near future. The boundary conditions could certainly also benefit from
some more elaborated methods like a sponge layer and non alteration of
the boundary value by the filtering process (not yet implemented).

We think that the most physical satisfactory way to get rid of the
Gibbs phenomenon related to the thin current sheet would be to
alleviate the ideal MHD approximation by including particle inertia.
Our code is sufficiently versatile and can be straightforwardly
extended to solve the full set of MHD equations. This would bring us
closer to reality in the striped structure. Moreover, the magnetic
field originates from a compact object of size only twice its
Schwarzschild radius. Therefore, we expect not only magnetospheric
currents distortion of the magnetic topology, but also general
relativistic effects, especially important in the vicinity of the
stellar surface and very useful for elucidating the precise geometry
of the polar caps. These perturbations due to curved space and
dragging of inertial frames, should also be taken into account for an
accurate investigation of the emission region close to the magnetic
poles.

\bsp

\appendix

\onecolumn

\section{Scalar and vector spherical harmonics}
\label{app:HSV}

In this appendix, we expose in some details our pseudo-spectral
algorithm to compute approximate analytical solutions to the
force-free pulsar magnetosphere according to the time-dependent
Maxwell equations.

The heart of our scheme resides in the vector spherical harmonic
expansion of the two unknown vector fields, the electric and magnetic
fields. In the following paragraphs, we discuss in depth the properties
of this vector basis. But let us first quickly remind the scalar spherical
harmonics.

\subsection{Scalar spherical harmonics}

The scalar spherical harmonics are defined according to the associated
Legendre functions~$P_l^m$ by
\begin{eqnarray}
  \label{eq:Y_lm_definition}
  Y_{lm}(\vartheta) & = & \sqrt{ \frac{2\,l+1}{4\,\pi} \, 
    \frac{(l-m)!}{(l+m)!} } \, P_l^m(\cos\vartheta) \\
  Y_{lm}(\vartheta,\varphi) & = & Y_{lm}(\vartheta) \, e^{i\,m\,\varphi}
\end{eqnarray}
The square root factor imposes normalization of these functions.  They
are very valuable functions because eigenfunctions of the Laplacian
operator in spherical coordinates such that
\begin{equation}
  \label{eq:Laplacien_Y_lm}
  \Delta Y_{lm}(\vartheta,\varphi) = - \frac{l\,(l+1)}{r^2} \, 
  Y_{lm}(\vartheta,\varphi)
\end{equation}
A useful symmetry property is
\begin{equation}
  Y_{l,-m}(\vartheta,\varphi) = (-1)^m \, Y_{lm}^*(\vartheta,\varphi)
\end{equation}
Note that the $Y_{lm}$ form an orthonormal basis on the
two-dimensional sphere.  Any smooth and regular scalar field can be
expanded on this sphere thanks to the spherical harmonics. As
explained in the next paragraph, this statement can be generalized to
a vector field defined on the sphere.

For completeness, we give the first few~$ Y_{lm}$ up to $m=3$ 
\begin{eqnarray*}
  Y_{00} & = & \frac{1}{2 \sqrt{\pi }} \\
  Y_{10} & = & \frac{1}{2} \sqrt{\frac{3}{\pi }} \cos[\vartheta ] \\
  Y_{11} & = & -\frac{1}{2} e^{i \varphi } \sqrt{\frac{3}{2 \pi }} \sin[\vartheta ] \\
  Y_{20} & = & \frac{1}{4} \sqrt{\frac{5}{\pi }} \left(-1+3 \cos[\vartheta ]^2\right) \\
  Y_{21} & = & -\frac{1}{2} e^{i \varphi } \sqrt{\frac{15}{2 \pi }} \cos[\vartheta
  ] \sin[\vartheta ] \\
  Y_{22} & = & \frac{1}{4} e^{2 i \varphi } \sqrt{\frac{15}{2 \pi }} \sin[\vartheta ]^2  \\
  Y_{30} & = & \frac{1}{4} \sqrt{\frac{7}{\pi }} 
  \left(-3 \cos[\vartheta ]+5 \cos[\vartheta ]^3\right) \\
  Y_{31} & = & -\frac{1}{8} e^{i \varphi } \sqrt{\frac{21}{\pi
    }} \left(-1+5 \cos[\vartheta ]^2\right) \sin[\vartheta ] \\
  Y_{32} & = & \frac{1}{4} e^{2 i \varphi } \sqrt{\frac{105}{2 \pi }} 
  \cos[\vartheta ] \sin[\vartheta]^2 \\
  Y_{33} & = & -\frac{1}{8} e^{3 i \varphi } \sqrt{\frac{35}{\pi }} \sin[\vartheta ]^3
\end{eqnarray*}

\subsection{Vector spherical harmonics}

Vector spherical harmonics (VSH) are often used in atomic physics to
compute electronic level in atoms. Quantum physicists therefore prefer
to work with a normalization of the vectors involving the complex
number~$i$. Here we adopt a slightly different definition of these
vectors by removing this unessential constant factor~$i$. More
precisely, the three sets of vector spherical harmonics we use are
defined by
\begin{eqnarray}
  \label{eq:Ylm_vect_def}
  \mathbf{Y}_{lm}    & = & Y_{lm} \, \mathbf{e}_{\rm r} \\
  \label{eq:Psilm_vect_def}
  \mathbf{\Psi}_{lm} & = & \frac{r}{\sqrt{l\,(l+1)}} \, \mathbf{\nabla} Y_{lm} \\
  \label{eq:Philm_vect_def}
  \mathbf{\Phi}_{lm} & = & \frac{\mathbf{r}}{\sqrt{l\,(l+1)}} \, \wedge \mathbf{\nabla} Y_{lm}
\end{eqnarray}
Any smooth three-dimensional vector field~$\mathbf{E}$ admits an
expansion onto these vectors according to
\begin{equation}
  \label{eq:Decomposition_HSV_general}
  \mathbf{E}(r,\vartheta,\varphi) = \sum_{l=0}^\infty\sum_{m=-l}^l
  \left(E^r_{lm}(r)\mathbf{Y}_{lm}+E^{(1)}_{lm}(r)\mathbf{\Psi}_{lm}+
    E^{(2)}_{lm}(r)\mathbf{\Phi}_{lm}\right)
\end{equation}
In order to avoid complex numbers, it is sometimes useful to employ
instead the trigonometric representation like for instance
\begin{eqnarray}
  \label{eq:Ylmcs_vect_def}
  \mathbf{Y}_{lm}^{c/s} & = & Y_{lm}(\vartheta) \, \begin{pmatrix} \cos m\,\varphi \\ \sin m\,\varphi \end{pmatrix} \, \mathbf{e}_{\rm r}
  %\mathbf{\Psi}_{lm}^{c/s} & = & \frac{r}{\sqrt{l\,(l+1)}} \, \mathbf{\nabla} Y_{lm}(\vartheta)  \, \begin{pmatrix} \cos m\,\varphi \\ \sin m\,\varphi \end{pmatrix} \\
  %\mathbf{\Phi}_{lm}^{c/s} & = & \frac{\mathbf{r}}{\sqrt{l\,(l+1)}} \, \wedge \mathbf{\nabla} Y_{lm}(\vartheta) \, \begin{pmatrix} \cos m\,\varphi \\ \sin m\,\varphi \end{pmatrix}
\end{eqnarray}
and similar expressions for $\{\mathbf{\Psi}_{lm}^{c/s},
\mathbf{\Phi}_{lm}^{c/s}\}$. 

To link our definition to other authors, note that in quantum
mechanics the~$\mathbf{\Phi}_{lm}$ are named~$\mathbf{X}_{lm}$ and
defined by
\begin{equation}
  \label{eq:Xlm_vect_def}
  \mathbf{X}_{lm} = \frac{-i}{\sqrt{l\,(l+1)}} \, \mathbf{r} \wedge \mathbf{\nabla} Y_{lm} = -i \, \mathbf{\Phi}_{lm}
\end{equation}

\subsection{Properties}

The vector spherical harmonics share some useful properties with
respect to their spatial derivatives. First, assume a 3D scalar field
expanded onto the scalar spherical harmonics such that
\begin{eqnarray}
  \label{eq:MultipoleScalaire}
  \phi(r,\vartheta,\varphi) & = & \sum_{l=0}^\infty \sum_{m=-l}^l \phi_{lm}(r) \, Y_{lm}(\vartheta,\varphi)
\end{eqnarray}
Then, its gradient expanded onto the VSH becomes
\begin{equation}
  \label{eq:Gradient}
  \nabla\phi = \sum_{l=0}^\infty \sum_{m=-l}^l 
  \left(\frac{\partial\phi_{lm}}{\partial r} \mathbf{Y}_{lm} +
    \frac{\sqrt{l\,(l+1)}}{r} \, \phi_{lm} \, \mathbf{\Psi}_{lm}\right) 
\end{equation}
The action of the same gradient on the VSH gives the divergence of any
vector field~$\mathbf{E}$ as
\begin{equation}
\nabla\cdot\mathbf{E} = \sum_{l=0}^\infty \sum_{m=-l}^l 
\left( \frac{1}{r^2} \, \frac{\partial}{\partial r} ( r^2 \, E^r_{lm} ) - 
  \frac{\sqrt{l(l+1)}}{r}E^{(1)}_{lm}\right)Y_{lm}
\end{equation}
and for the curl
\begin{eqnarray}
\nabla \wedge \mathbf{E} & = & \sum_{l=0}^\infty \sum_{m=-l}^l \left[ - \frac{\sqrt{l(l+1)}}{r} \, E^{(2)}_{lm} \, \mathbf{Y}_{lm} - \frac{1}{r} \, \frac{\partial}{\partial r} (r\,E^{(2)}_{lm}) \mathbf{\Psi}_{lm} + \left(- \frac{\sqrt{l(l+1)}}{r} \, E^r_{lm} + \frac{1}{r} \, \frac{\partial}{\partial r}(r\,E^{(1)}_{lm}) \right) \, \mathbf{\Phi}_{lm} \right]
\end{eqnarray}
For each component taken individually, we find for the divergence
\begin{eqnarray}
  \label{eq:Div}
  \nabla \cdot \left(f(r) \, \mathbf{Y}_{lm} \right) & = &
  \frac{1}{r^2} \frac{\partial}{\partial r} \left( r^2 \, f \right) \, Y_{lm} \\
  \nabla \cdot \left(f(r) \, \mathbf{\Psi}_{lm}\right) & = &
  - \frac{\sqrt{l(l+1)}}{r} \, f \, Y_{lm} \\
  \nabla\cdot\left(f(r)\mathbf{\Phi}_{lm}\right) & = & 0
\end{eqnarray}
and for the curl
\begin{eqnarray}
  \label{eq:Rot}
  \nabla \times \left(f(r) \, \mathbf{Y}_{lm}\right) & =- & 
  \frac{\sqrt{l\,(l+1)}}{r} \, f \, \mathbf{\Phi}_{lm} \\
  \nabla\times\left(f(r)\mathbf{\Psi}_{lm}\right) & = &
  \frac{1}{r} \, \frac{\partial}{\partial r}(r\,f) \mathbf{\Phi}_{lm} \\
  \nabla\times\left(f(r)\mathbf{\Phi}_{lm}\right) & = & - \frac{\sqrt{l(l+1)}}{r} \, f \, \mathbf{Y}_{lm} - \frac{1}{r} \, \frac{\partial}{\partial r}(r\,f) \, \mathbf{\Psi}_{lm}
\end{eqnarray}
Finally, define the radial differential operator~$\mathcal{D}_l$ by
\begin{equation}
  \label{eq:MatchalD}
  \mathcal{D}_l = \frac{1}{r^2} \, \frac{d}{dr} \left( r^2 \, \frac{d}{dr}\right)
  - \frac{l(l+1)}{r^2}
\end{equation}
The VSH noted $\mathbf{\Phi}_{lm}$ are eigenvectors in the linear
algebra meaning, of the vector Laplacian operator~$\Delta$. Indeed, it
is straightforward to show that
\begin{equation}
  \label{eq:nabl}
  \Delta [ f(r) \, \mathbf{\Phi}_{lm} ] = \mathcal{D}_l[f] \, \mathbf{\Phi}_{lm}
\end{equation}
It is thus an extension of the properties of the scalar spherical
harmonics to the realm of 3D vectors.

\subsection{Expansion of a vector field onto VSH}

From the above discussion, the components of any vector field can be
computed according to the three underlying equalities
\begin{eqnarray}
 E_r & = & \sum_{l,m} E_{lm}^r \, Y_{lm} \\
 \divg_{\vartheta,\varphi} \vec E & = & \sum_{l,m} - \frac{\sqrt{l(l+1)}}{r} \, E_{lm}^{(1)} \, Y_{lm} \\
 \rot \vec E \cdot \er & = & \sum_{l,m} - \frac{\sqrt{l(l+1)}}{r} \, E_{lm}^{(2)} \, Y_{lm}
\end{eqnarray}
where $\divg_{\vartheta,\varphi} \vec E$ means taking only the angular
part of the divergence.  More explicitly, from the definition of the
differential operators, we get
\begin{eqnarray}
 E_r & = & \sum_{l,m} E_{lm}^r \, Y_{lm} \\
 \frac{1}{\sin\vartheta} \, \partial_\vartheta (\sin\vartheta \, E_\vartheta ) +  \frac{1}{\sin\vartheta} \, \partial_\varphi \, E_\varphi & = & \sum_{l,m} - \sqrt{l(l+1)} \, E_{lm}^{(1)} \, Y_{lm} \\
 \frac{1}{\sin\vartheta} \, \partial_\vartheta (\sin\vartheta \, E_\varphi ) -  \frac{1}{\sin\vartheta} \, \partial_\varphi \, E_\vartheta & = & \sum_{l,m} - \sqrt{l(l+1)} \, E_{lm}^{(2)} \, Y_{lm}
\end{eqnarray}
Finding the components of $\mathbf{E}$ is therefore equivalent to finding
the expansion coefficients of three scalar fields onto the scalar
spherical harmonics. This procedure works for any vector
field. However, the magnetic field being divergenceless, only two of
the three components are independent. It is therefore judicious to
deal properly with those kind of fields by analytically enforcing the
condition on the divergence as explained below.

\subsection{Expansion of a divergenceless vector field}

Any divergenceless vector field is efficiently developed onto the
vector spherical harmonic \textit{orthonormal basis}. This will be the
case for the magnetic field in our algorithm.

Assume that the vector field~$\mathbf{V}$ is divergenceless. It is helpful to
introduce two scalar functions $f_{lm}(r,t)$ and $g_{lm}(r,t)$ such
that the decomposition immediately implies the property of
divergenceless field. This is achieved by writing
\begin{equation}
  \label{eq:Decomposition_HSV_div_0}
  \mathbf{V}(r,\vartheta,\varphi,t) = \sum_{l=1}^\infty\sum_{m=-l}^l
  \left( \rot [f_{lm}(r,t) \, \mathbf{\Phi}_{lm}] + 
    g_{lm}(r,t) \, \mathbf{\Phi}_{lm} \right)
\end{equation}
This expression automatically and \textit{analytically} enforces the
condition $\divg \mathbf{V} = 0$.  Let us quickly draw the way to
compute these functions.  The transformation from the spherical
components to the functions $(f_{lm},g_{lm})$ is given by
\begin{eqnarray}
 \mathbf{V} \cdot \er & = & \sum_{l=1}^\infty\sum_{m=-l}^l - \frac{\sqrt{l\,(l+1)}}{r} \, f_{lm} \, Y_{lm} \\
 (\rot \mathbf{V}) \cdot \er & = & \sum_{l=1}^\infty\sum_{m=-l}^l - \frac{\sqrt{l\,(l+1)}}{r} \, g_{lm} \, Y_{lm}
\end{eqnarray}
Thus, it is sufficient to expand again the radial component of the
vector and its curl onto scalar spherical harmonics. We get
\begin{eqnarray}
 r \, V_r & = & \sum_{l,m} - \sqrt{l\,(l+1)} \, f_{lm} \, Y_{lm} \\
 \frac{1}{\sin\vartheta} \, \partial_\vartheta (\sin\vartheta \, V_\varphi ) -  \frac{1}{\sin\vartheta} \, \partial_\varphi \, V_\vartheta & = & \sum_{l,m} - \sqrt{l(l+1)} \, g_{lm} \, Y_{lm}
\end{eqnarray}
The functions $\{f_{lm},g_{lm}\}$ are related to the general expansion
Eq.~(\ref{eq:Decomposition_HSV_general}) by
\begin{eqnarray}
 V_{lm}^r & = & - \frac{\sqrt{l\,(l+1)}}{r} \, f_{lm} \\
 V_{lm}^{(1)} & = & - \frac{1}{r} \, \partial_r(r\,f_{lm}) \\
 V_{lm}^{(2)} & = & g_{lm}
\end{eqnarray}

\subsection{The first few VSH}

Here also, for completeness, we give the first few VSH which can be
useful for analytical calculations.

The VSH~$\mathbf{\Psi}_{lm}$ are given in the orthonormal basis
$\{\er, \etheta, \ephi\}$ by
\begin{eqnarray*}
  \mathbf{\Psi}_{10} & = & \{0,\; 
  -\frac{1}{2} \sqrt{\frac{3}{2\,\pi }} \sin[\vartheta ],\; 0 \} \\
  \mathbf{\Psi}_{11} & = & \{0,\;
  -\frac{1}{2} e^{i \varphi } \sqrt{\frac{3}{4 \pi }} \cos[\vartheta ],\;
  -\frac{1}{2} i e^{i \varphi } \sqrt{\frac{3}{4 \pi }} \} \\
  \mathbf{\Psi}_{20} & = & \{0,\;
  -\frac{3}{2} \sqrt{\frac{5}{6\pi }} \cos[\vartheta ] \sin[\vartheta ],\;
  0 \} \\
  \mathbf{\Psi}_{21} & = & \{0,\;
  -\frac{1}{2} e^{i \varphi } \sqrt{\frac{15}{12 \pi }} \cos[\vartheta ]^2 + 
  \frac{1}{2} e^{i \varphi } \sqrt{\frac{15}{12 \pi }} \sin[\vartheta]^2,\;
  -\frac{1}{2} i e^{i \varphi } \sqrt{\frac{15}{12 \pi }} \cos[\vartheta ] \} \\
  \mathbf{\Psi}_{22} & = & \{0,\;
  \frac{1}{2} e^{2 i \varphi } \sqrt{\frac{15}{12 \pi }} \cos[\vartheta ] \sin[\vartheta ],\;
  \frac{1}{2} i e^{2 i \varphi } \sqrt{\frac{15}{12 \pi }} \sin[\vartheta ] \}\\
  \mathbf{\Psi}_{30} & = & \{0,\;
  \frac{1}{4} \sqrt{\frac{7}{12\pi }} \left(3 \sin[\vartheta ] - 
    15 \cos[\vartheta ]^2 \sin[\vartheta ]\right),\; 0 \} \\
  \mathbf{\Psi}_{31} & = & \{0,\;
  -\frac{1}{8} e^{i \varphi } \sqrt{\frac{21}{12\pi }} \cos[\vartheta ] 
  \left(-1+5 \cos[\vartheta ]^2\right)+\frac{5}{4} e^{i \varphi } \sqrt{\frac{21}{12\pi
    }} \cos[\vartheta ] \sin[\vartheta ]^2,\;
  -\frac{1}{8} i e^{i \varphi } \sqrt{\frac{21}{12\pi }} \left(-1+5 \cos[\vartheta ]^2\right) \} \\
  \mathbf{\Psi}_{32} & = & \{0,\;
  \frac{1}{2} e^{2 i \varphi } \sqrt{\frac{105}{24 \pi }} \cos[\vartheta ]^2 \sin[\vartheta] - 
  \frac{1}{4} e^{2 i \varphi } \sqrt{\frac{105}{24 \pi }} \sin[\vartheta ]^3,\;
  \frac{1}{2} i e^{2 i \varphi } \sqrt{\frac{105}{24 \pi }} \cos[\vartheta ] \sin[\vartheta ] \} \\
  \mathbf{\Psi}_{33} & = & \{0,\;
  -\frac{3}{8} e^{3 i \varphi } \sqrt{\frac{35}{12\pi }} \cos[\vartheta ] \sin[\vartheta ]^2,\;
  -\frac{3}{8} i e^{3 i \varphi } \sqrt{\frac{35}{12\pi }} \sin[\vartheta ]^2 \}
\end{eqnarray*}
whereas the~$\mathbf{\Phi}_{lm}$ are expressed as
\begin{eqnarray*}
  \mathbf{\Phi}_{10} & = & \{0,\;0,\;
  -\frac{1}{2} \sqrt{\frac{3}{2\pi }} \sin[\vartheta ] \} \\
  \mathbf{\Phi}_{11} & = & \{0,\;
  \frac{1}{2} i e^{i \varphi } \sqrt{\frac{3}{4 \pi }},\;
  -\frac{1}{2} e^{i \varphi } \sqrt{\frac{3}{4 \pi }} \cos[\vartheta ] \} \\
  \mathbf{\Phi}_{20} & = & \{0,\;0,\;
  -\frac{3}{2} \sqrt{\frac{5}{6\pi }} \cos[\vartheta ] \sin[\vartheta ] \} \\
  \mathbf{\Phi}_{21} & = & \{0,\;
  \frac{1}{2} i e^{i \varphi } \sqrt{\frac{15}{12 \pi }} \cos[\vartheta ],\;
  -\frac{1}{2} e^{i \varphi } \sqrt{\frac{15}{12 \pi }} \cos[\vartheta ]^2 + 
  \frac{1}{2} e^{i \varphi } \sqrt{\frac{15}{12 \pi }} \sin[\vartheta]^2 \} \\
  \mathbf{\Phi}_{22} & = & \{0,\;
  -\frac{1}{2} i e^{2 i \varphi } \sqrt{\frac{15}{12 \pi }} \sin[\vartheta ],\;
  \frac{1}{2} e^{2 i \varphi } \sqrt{\frac{15}{12 \pi }} \cos[\vartheta ] \sin[\vartheta ]\} \\
  \mathbf{\Phi}_{30} & = & \{0,\;0,\;
  \frac{3}{4} \sqrt{\frac{7}{12\pi }} \sin[\vartheta ] -
  \frac{15}{4} \sqrt{\frac{7}{12\pi }} \cos[\vartheta ]^2 \sin[\vartheta ] \} \\
  \mathbf{\Phi}_{31} & = & \{0,\;
  \frac{1}{8} i e^{i \varphi } \sqrt{\frac{21}{12\pi }} \left(-1+5 \cos[\vartheta ]^2\right),\;
  \frac{1}{8} e^{i \varphi } \sqrt{\frac{21}{12\pi }} \cos[\vartheta ]-\frac{5}{8} e^{i \varphi } \sqrt{\frac{21}{12\pi }} \cos[\vartheta ]^3+\frac{5}{4}
  e^{i \varphi } \sqrt{\frac{21}{12\pi }} \cos[\vartheta ] \sin[\vartheta ]^2 \} \\
  \mathbf{\Phi}_{32} & = & \{0,\;
  -\frac{1}{2} i e^{2 i \varphi } \sqrt{\frac{105}{24 \pi }} \cos[\vartheta ] \sin[\vartheta ],\;
  \frac{1}{2} e^{2 i \varphi } \sqrt{\frac{105}{24 \pi }} 
  \cos[\vartheta ]^2 \sin[\vartheta ]-\frac{1}{4} e^{2 i \varphi } \sqrt{\frac{105}{24
      \pi }} \sin[\vartheta ]^3 \} \\
  \mathbf{\Phi}_{33} & = & \{0,\;
  \frac{3}{8} i e^{3 i \varphi } \sqrt{\frac{35}{12\pi }} \sin[\vartheta ]^2,\;
  -\frac{3}{8} e^{3 i \varphi } \sqrt{\frac{35}{12\pi }} \cos[\vartheta ] \sin[\vartheta ]^2 \}
\end{eqnarray*}

\section{Deutsch field}
\label{app:Deutsch}

For comparison with numerical simulations, we recall the exact complex
expressions for the Deutsch vacuum field solution in the general
oblique case. The magnetic field, dipolar close to the neutron star,
is given by
\begin{eqnarray}
  \label{eq:DeutschB}
  B_r(\mathbf{r},t) & = & 2 \, B_* \, \left[ \frac{R_*^3}{r^3} \, \cos\chi \, \cos\vartheta + 
    \frac{R_*}{r} \, \frac{h^{(1)}_1(k\,r)}{h^{(1)}_1(k\,R_*)} \, 
    \sin\chi \, \sin \vartheta \, e^{i\,\psi} \right] \\
  B_\vartheta(\mathbf{r},t) & = & B_* \, \left[ \frac{R_*^3}{r^3} \, \cos\chi \, \sin\vartheta + \left( \frac{R_*}{r} \, \frac{\frac{d}{dr} \left( r \, h^{(1)}_1(k\,r) \right)}{h^{(1)}_1(k\,R_*)} + \frac{R_*^2}{\rlight^2} \, \frac{h^{(1)}_2(k\,r)}{\frac{d}{dr} \left( r \, h^{(1)}_2(k\,r) \right) |_{R_*}} \right) \, 
    \sin\chi \, \cos \vartheta \, e^{i\,\psi} \right] \\
  B_\varphi(\mathbf{r},t) & = & B_* \, \left[ \frac{R_*}{r} \, 
    \frac{\frac{d}{dr} ( r \, h^{(1)}_1(k\,r) )}{h^{(1)}_1(k\,R_*)} \, 
    + \frac{R_*^2}{\rlight^2}
    \frac{h^{(1)}_2(k\,r)}{\frac{d}{dr} \left( r \, h^{(1)}_2(k\,r) \right) |_{R_*}}
    \, \cos 2\,\vartheta \right] \, i \, \sin\chi \, \, e^{i\,\psi}
\end{eqnarray}
The induced electric field is then
\begin{eqnarray}
  \label{eq:DeutschE}
  E_r(\mathbf{r},t) & = & \Omega_* \, B_* \, R_* \, 
  \left[ \left( \frac{2}{3} - \frac{R_*^2}{r^2} ( 3 \, \cos^2\vartheta - 1 ) \right) \
    \, \frac{R_*^2}{r^2} \, \cos\chi + 3 \, \sin\chi\, \sin 2\,\vartheta \, e^{i\,\psi}  \,
    \frac{R_*}{r} \, \frac{ h^{(1)}_2(k\,r)}
    {\frac{d}{dr} \left( r \, h^{(1)}_2(k\,r) \right) |_{R_*}} \right] \\
  E_\vartheta(\mathbf{r},t) & = & \Omega_* \, B_* \, R_* \, 
  \left[ - \frac{R_*^4}{r^4} \sin 2\,\vartheta \, \cos\chi + \sin\chi\, e^{i\,\psi}  \, \left(
      \frac{R_*}{r} \, \frac{\frac{d}{dr} \left( r \, h^{(1)}_2(k\,r) \right)}
      {\frac{d}{dr} \left( r \, h^{(1)}_2(k\,r) \right)|_{R_*}} \, \cos 2\,\vartheta -
      \frac{h^{(1)}_1(k\,r)}{h^{(1)}_1(k\,R_*)} \right) \right] \\
  E_\varphi(\mathbf{r},t) & = & \Omega_* \, B_* \, R_* \, \left[ \frac{R_*}{r} \, 
    \frac{\frac{d}{dr} \left( r \, h^{(1)}_2(k\,r) \right)}
    {\frac{d}{dr} \left( r \, h^{(1)}_2(k\,r) \right)|_{R_*}} -
    \frac{h^{(1)}_1(k\,r)}{h^{(1)}_1(k\,R_*)} \right] \, i \sin\chi \, \cos\vartheta \, e^{i\,\psi}
\end{eqnarray}
$k=1/\rlight$ is the wavenumber and $h^{(1)}_l$ are the spherical
Hankel functions of order~$l$ satisfying the outgoing wave conditions,
see for instance \cite{1995mmp..book.....A}.  The physical solution is
found by taking the real parts of each component, it encompasses a
linear combination of the vacuum aligned dipole field and the vacuum
orthogonal rotator with respective weights $\cos\chi$ and
$\sin\chi$. To complete the solution, we add a monopolar electric
field contribution due to the stellar surface charge such that
\begin{equation}
 E_r^{\rm mono} = \frac{Q_*-Q_{\rm c}}{4\,\pi\,\varepsilon_0\,r^2}
\end{equation} 
where $Q_*$ is the total electric charge of the star and
\begin{equation}
  \label{eq:QC}
  Q_{\rm c} = \frac{8\,\pi}{3}\,\varepsilon_0 \, \Omega_* \, B_* \, R_*^3 \, \cos\chi
\end{equation}
is the central point charge inside the star.  The above expressions
reduce to the static oblique dipole for small distances $r\ll \rlight$
(static zone).

\label{lastpage}

\end{document}